\documentclass[12pt]{iopart} % for review and submission
\usepackage{graphicx}  % needed for figures
\usepackage{dcolumn}   % needed for some tables
\usepackage{bm}        % for math
\usepackage{amssymb}   % for math
\usepackage{color}
\usepackage{ulem}
\newcommand{\eqref}[1]{(\ref{#1})}

\begin{document}

\title{Entropy production for coarse-grained dynamics}
\author{D M Busiello$^{1,2}$, J Hidalgo$^1$ and
A Maritan$^1$}
\address{$^1$ Dipartimento di Fisica `G. Galilei', INFN, Universit\'a di Padova, Via Marzolo 8, 35131 Padova, Italy}
\address{$^2$ Laboratory of Statistical Biophysics, SB ITP, Ecole Polytechnique F\'ed\'erale de Lausanne (EPFL), CH-1015, Lausanne, Switzerland}
\ead{daniel.busiello@epfl.ch}

\begin{abstract}
Systems out of equilibrium exhibit a net production of entropy. We study the dynamics of a stochastic system represented by a Master Equation that can be modeled by a Fokker-Planck equation in a coarse-grained, mesoscopic description. We show that the corresponding coarse-grained entropy production contains information on microscopic currents that are not captured by the Fokker-Planck equation and thus cannot be deduced from it. {\color{black} We study a discrete-state and a continuous-state system, deriving in both the cases an analytical expression for the coarse-graining corrections to the entropy production. This result elucidates the limits in which there is no loss of information in passing from a Master Equation to a Fokker-Planck equation describing the same system. Our results are amenable of experimental verification, which could help to infer some information about the underlying microscopic processes.}
\end{abstract}

\maketitle

{\color{black}\section{Introduction}}

Any physical system, and its characterizing processes, can be depicted by making use of different levels of description. Considering a microscopic spatial and temporal resolution, any evolution will appear purely reversible in time. Since most of the details of a system are usually unknown, they are neglected \textit{a-priori}, thus requiring a mesoscopic description in terms of random variables and probabilities. The theory of stochastic thermodynamics relies on this assumption, i.e. on a temporal and spatial `coarse-graining' \cite{espos}. Furthermore, within the possible mesoscopic descriptions, different levels of coarse-graining are allowed, and all the physical observables could be somehow affected by the information we are unaware of or deliberately ignored \textit{a-priori}. Quantifying the influence of the coarse-graining on our prediction of the physical properties of a system is a long-standing problem, addressed by countless works in literature \cite{hasel, peles, pigo, santi, gomez, nicol}.

It is known \cite{espos} how the entropy balance is affected by performing a coarse-graining on the system `microstates'. The limit of instantaneous equilibration of the internal microscopic states makes the mathematical form of the theory independent of the level of description. Remarkably, this unravels the key assumption of the stochastic thermodynamics, that is the internal structure of each state may evolve in time, but always remaining at equilibrium. In  \cite{espos} the effect of neglecting information is investigated in a Markovian discrete-state dynamics, which is one of the possible ways to describe a stochastic system.

Among all the possible quantities that can be estimated in a system out of equilibrium, in this {\color{black}work} we focus on the entropy production, a fingerprint of non-equilibrium conditions. {\color{black} Recently, its crucial role in the outmost thermodynamic uncertainty relations \cite{dechant,barato} has been pointed out, along with the possibility to use the entropy production as a possible quantification of the non-equilibrium activity of a biological system \cite{horow}. It is also a fundamental quantity involved in various fluctuation theorems \cite{sekimoto, lebo, gall, maes, kurch, jarz, crooks, seif2}, whose theoretical relevance has stimulated several experimental confirmations in the field of stochastic thermodynamics \cite{evans, coll, cil}.} Moreover, the production of entropy has a leading role in building efficient engines \cite{how, brown}, since it can be understood as the `cost' of performing a given task. For all these reasons it has been widely investigated both in discrete \cite{schn, busie, raz} and continuous systems \cite{seif3, celani, pigo2, busie2}.

We consider a system with a finite number, $N$, of accessible states whose dynamics is described by a Master Equation (ME) of the form:
\begin{equation}
\dot{P}_i(t) = \sum_{j=1}^N \left( W_{ij} P_j(t) - W_{ji} P_i(t) \right)
\label{discME}
\end{equation}
where $W_{ij}$ is the transition rate from the state $j$ to the state $i$ and $P_i(t)$ is the probability to be in the state $i$ at time $t$. Following Schnakenberg's formulation \cite{schn}, the (average) entropy production is
\begin{equation}
\dot{S}_\mathrm{ME}(t) = \sum_{ij} W_{ij} P_j(t) \log\left(\frac{W_{ij} P_j(t)}{W_{ji} P_i(t)}\right),
\label{schn}
\end{equation}
where the sum is performed over all non-zero transition rates (it is assumed that $W_{ij}>0$ implies $W_{ji}>0$). Eq. \eqref{schn} was originally motivated from an information theory approach \cite{schn,shan}, but it is thermodynamically consistent, as pointed out in \cite{tome, busie}. In what follows we refer to Eq. \eqref{schn} as the \textit{microscopic} entropy production.

{\color{black} The entropy production is intimately connected to the information theory \cite{landauer,bennett}. Several experiments have been performed in this direction \cite{ciliberto,beck,gavrilov}, evidencing the physical meaning of the mathematical backbone on which this and previous works strongly relies.}

Stochastic systems, under suitable conditions, can be also described in terms of continuous variables by means of a diffusive equation.
The standard approach \cite{gardiner} consists of introducing a new variable $x=i\Delta x$, that represents, for example, the spatial position of a particle in the state $i$, which becomes continuous in the limit $\Delta x\rightarrow 0$. By performing the Kramers-Moyal expansion on Eq. \eqref{discME} \cite{gardiner}, this procedure leads to the Fokker-Planck equation (FPE) \cite{gardiner, redner}:
\begin{eqnarray}
\dot{P}(x,t) &=& -\partial_x \left[ A(x) P(x,t)  -  \partial_x \left(D(x) P(x,t)\right) \right] \nonumber\\
 &\equiv& -\partial_x [ J(x,t) ].
\label{FPE}
\end{eqnarray}
where $P(x,t)=P_i(t)/\Delta x$ represents the probability density function to be in the state $x$ at time $t$, $A(x)\equiv A(i\Delta x) = \sum_j (j-i)\Delta x W_{ji}$ the drift and $D(x)\equiv \frac{1}{2}\sum_j ((j-i)\Delta x)^2 W_{ji}$ the diffusion coefficient, in the limit $\Delta x\rightarrow 0$. 
% where $J(x,t)=\left(A(x)-D(x)\partial_x \right)P(x,t)$ is the probability current.
This approach relies on the assumption that all the `pseudo-moments' of the transition rates of order higher than $2$ vanish when $\Delta x$ approaches $0$ \cite{note}. It is important to notice that the dynamics represented by Eq. \eqref{FPE} belongs to a different level of description with respect to the discrete-state dynamics, Eq. (\ref{discME}), and all the relevant information are now encoded in the coefficients $A(x)$ and $D(x)$.

In \cite{seif3}, Seifert calculated the mean entropy production for systems described by a FPE starting from the entropy associated with each possible trajectory, leading to the following formula:
\begin{equation}
\dot{S}_\mathrm{FP}(t) = \int \frac{J(x,t)^2}{D(x) P(x,t)} dx.
\label{seifert}
\end{equation}

In this {\color{black}work}, we address the basic question of how equations \eqref{schn} and \eqref{seifert} are related.
The former is derived within a framework considering discrete states systems, whereas the latter arises directly in the continuum limit, where many microscopic details are ignored, {\color{black} i.e. after a suitable coarse-graining on the dynamics}. Since both formulas refer to the same quantity at two different levels of description, we naively expect that one can be obtained from the other.
As we will show, this is true only for a specific choice of the transition rates. However, in general, Eq. \eqref{seifert} does not fully capture the contribution to the entropy production stemming from the microscopic currents, which do not enter explicitly in the FPE.
% In other words, the continuum limit of the entropy production is affected by microscopic details not entering in the Fokker-Planck equation.

{\color{black}\section{Discrete-state systems}}

\begin{figure}[t]
\centering
\includegraphics[width=0.6\columnwidth]{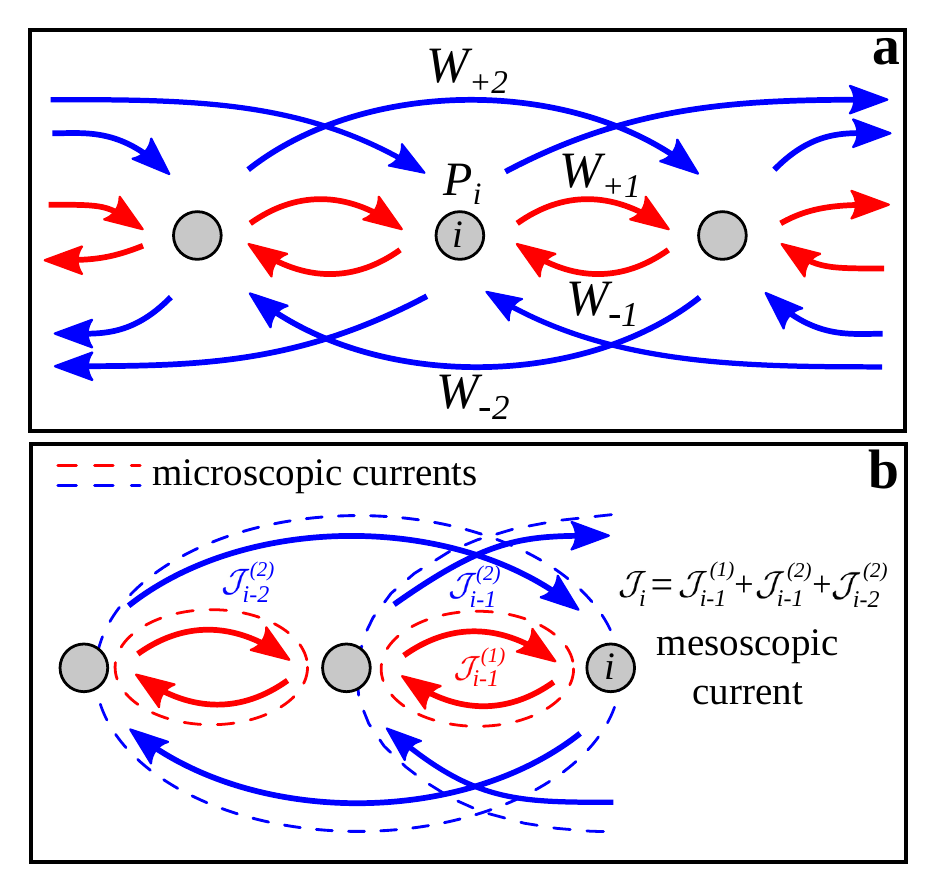}
\caption{ 
\textit{Panel a} - The microscopic dynamics of a $n$-step random walk is sketched: red and blue arrows indicate jumps to the right and left of size 1 and 2 with transition rates $W_{\pm1}$ and $W_{\pm2}$, respectively. \textit{Panel b} - Microscopic currents at each node $i$ can be associated with each jump size, $\mathcal{J}_i^{(k)}$, where $k=1,2$. The coarse-grained current, $J_i$, can be calculated considering all currents passing through a given node. This is the current appearing in the FPE.
}
\label{fig:1}
\end{figure}

As an illustration of the idea, we first consider a simple model of a one-dimensional random walk where a particle can jump in both directions with different step lengths $k=1,2,...,n$ at any time (for simplicity in the formulation we skip the length scale at this point), as sketched in Fig. \ref{fig:1} {\color{black} \cite{notefoot1}}. Jump rates are:
\begin{equation}
W_{ij} =\left\{
\begin{array}{ll}
		     W_{\pm k} \delta_{j,i\pm k},&\quad k = 1,...,n \\
                     0&\quad\mathrm{otherwise.}
\end{array}\right.
\end{equation}

{\color{black}The ME for this process can be written in terms of the incoming and outgoing probability currents at each node, $\dot P_i(t) = \mathcal{J}^\mathrm{+}_i(t)-\mathcal{J}^\mathrm{-}_i(t)$, where:
\begin{eqnarray}
\mathcal{J}^\mathrm{+}_i(t) &=&  \sum_{k=1}^n \mathcal{J}^{(k)}_{i-k}(t),\\
\mathcal{J}^\mathrm{-}_i(t) &=& \sum_{k=1}^n \mathcal{J}^{(k)}_{i}(t)
\end{eqnarray}
with
\begin{equation}	
\mathcal{J}^{(k)}_i(t) = W_{+k} P_i(t) - W_{-k} P_{i+k}(t),\quad k=1,...n.
\label{jk}
\end{equation}
being the instantaneous current passing through the node $i$ at time $t$ due to jumps of size $k$.

\subsection{Probability currents and entropy production inequality}

The microscopic entropy production, as defined by Schnakenberg \cite{schn}, can be written in terms of the microscopic currents:
\begin{equation}
\dot{S}_\mathrm{ME} = -\sum_i \sum_{k=1}^n \mathcal{J}_i^{(k)}(t) \log\left( 1 - \frac{\mathcal{J}_i^{(k)}(t)}{W_{+k}P_{i}(t)} \right).
\label{SME}
\end{equation}}

A description in terms of a FPE, Eq. \eqref{FPE}, can be guaranteed if we take the transition rates as
\begin{equation}
W_{\pm k} =\Big(1 + \frac{\beta_k \pm \alpha_k}{2}\Delta x \Big)\frac{w_k}{\Delta x^2},
\label{Wpm}
\end{equation}
where $w_k\geq 0$ and $\beta_k\geq |\alpha_k|$ to ensure that $W_{\pm k} \geq 0$ for all $\Delta x$. In particular, Eq. \eqref{Wpm} leads to $A = \sum_{k=1}^n k\alpha_kw_k$ and $D = \sum_{k=1}^n D_k$, where $D_k = k^2w_k$ is the diffusion coefficient associated with the process involving only jumps of size $k$, and higher-order `pseudo-moments' of the transition rates vanish when $\Delta x\rightarrow 0$. Note that in this simple case both coefficients are independent of $x$ {\color{black}\cite{notefoot2}}.

The microscopic entropy production, {\color{black}Eq. \eqref{SME}}, in the continuum limit becomes:
\begin{equation}
\dot{S}_\mathrm{ME}^{\Delta x\rightarrow 0} = \sum_{k=1}^n \int dx \frac{\mathcal{J}^{(k)}(x,t)^2}{D_{k} P(x,t)}.
\label{SME0}
\end{equation}
where, {\color{black}by definition, the probability} current associated with the step of size $k$ {\color{black}in the continuum limit is}
\begin{eqnarray}
 \mathcal{J}^{(k)}(x,t) = kw_k\big(\alpha_k P(x,t)-k\partial_xP(x,t)\big)
 \label{currentk}
\end{eqnarray}

{\color{black}Alternatively, it is possible to write the ME in terms of a current accounting for \textit{all} the probability flux crossing a fictitious barrier located at node $i$ (see Fig. \ref{fig:1}):
\begin{equation}
J_i(t) = \sum_{k=1}^n \sum_{m=1}^k \mathcal{J}_{i-m}^{(k)}(t),
\end{equation}
so that $\dot P_i(t) = -(J_{i+1}(t)-J_i(t))$. In the continuum limit, the probability current $J_i(t)=\sum_{k=1}^n k \mathcal{J}^{(k)}_{i}(t) \rightarrow \sum_{k=1}^n \mathcal{J}^{(k)}(x,t)=J(x,t)$, which is the corresponding current appearing in the FPE.

Thus the Seifert's formula for the entropy production, Eq.\eqref{seifert}, gives:}
\begin{equation}
\dot{S}_{\mathrm{FP}} = \int dx \frac{\left( \sum_k \mathcal{J}^{(k)}(x,t) \right)^2}{\sum_k D_{k} P(x,t)},
%\label{Sineq}
\end{equation}
{\color{black}As a consequence of the Cauchy-Schwarz inequality, we get}
\begin{equation}
\dot{S}_{\mathrm{FP}}  \leq \dot{S}_{ME}^{\Delta x \rightarrow 0},
\label{Sineq}
\end{equation}
It is interesting to note that Eq. \eqref{SME0} corresponds to the sum of the mesoscopic entropy production, as in Eq. \eqref{seifert}, associated with each microscopic process, while the entropy production directly derived from a mesoscopic description involves an `integrated' current and diffusion coefficient, leading to the inequality in Eq. \eqref{Sineq}.

{\color{black}\subsection{Interpretation of the results and conditions for not having corrections}}

We conclude that Seifert's formula represents a lower bound for the production of entropy; instead, Eq.~\eqref{SME0} gives a more accurate value as it captures all microscopic currents hidden in the mesoscopic description, but contributing to the entropy production.

Intuitively, the discrepancy between the two formulas relies on having different `channels' through which the particle can move, jumping to distant locations without necessarily going through the intermediate points (see Fig. \ref{fig:1}). All the microscopic currents contribute to the production of entropy. When the system is coarse-grained they are simply added up and part of the information is lost if currents through different channels flow in opposite directions.

{\color{black}The inequality in Eq. \eqref{Sineq} is formally equivalent to the one derived in \cite{espos,threefaces}, for a system whose transitions occur due to the coupling to different baths. However, we point out that this latter is just a mathematical similarity, since the `channels' we refer to are just ensemble of transitions of different length that cannot be resolved within a diffusive description.}

Remarkably, notice that if $\mathcal{J}_{i}^{(k)}=0$, that is the microscopic detailed balance condition is satisfied, then also detailed balance holds in the corresponding Fokker-Planck description, i.e.  $J(x)=0$ \cite{gardiner}. However, the vice versa does not necessarily hold, that is an equilibrium in the continuum description does not necessarily implies that the underlying microscopic dynamics is also at equilibrium. 
%Indeed, if $\sum_{k=1}^n k\alpha_kw_k=0$ we obtain $\dot S_\mathrm{FP}^*=0$  whereas $\dot{S}_\mathrm{ME}^{\Delta x \rightarrow 0,*}>0$, as soon as $\alpha_k$ is not proportional to  $k$.  
In other words, the system seems at equilibrium in the continuum description, while there is not detailed balance at the microscopic level.

{\color{black}It is worth noting that in some conditions the Seifert's formula captures all the relevant information about the system, and no coarse-graining correction to the entropy production is needed. Independently of the details of the specific model, this happens when $k=1$, i.e. only transitions between nearest neighbors states are allowed. In other words, this condition is equivalent of having just one single `channel' through which the particle can jump. 

Another example of a physical system satisfying this constraint, in a two-dimensional space, is presented in \cite{gingrich}, where the formula for the entropy production is the generalization of the Seifert's one, as derived in \cite{pigo2}.}
\\

{\color{black}\subsection{A simple example}

The multi-step random walk becomes very simple to solve at stationarity if we impose periodic boundary conditions. The stationary solution of the Master Equation corresponds to the homogeneous state $P^*(x)=1/L$, where $L$ is the size of the system. Thus, the 
Seifert's formula for the entropy production simplifies to:
\begin{equation}
\dot{S}_\mathrm{FP}^*=\frac{(\sum_{k=1}^n k\alpha_kw_k)^2}{\sum_{k=1}^n k^2w_k}
\end{equation}
whereas the actual value for the entropy production can be found by taking the continuum limit of the microscopic entropy production:
\begin{eqnarray}
\dot{S}_\mathrm{ME}^{\Delta x \rightarrow 0,*} &=& \sum_{k=1}^n \alpha_k^2w_k= \dot{S}_\mathrm{FP}^* +\\ &+&\frac{\sum_{1\leq k< k'\leq n} w_kw_{k'}(k\alpha_{k'} -k'\alpha_k)^2}{\sum_{k=1}^n k^2w_k}.
\label{SMEranwalk}
\end{eqnarray}
Apart from the trivial case of $n=1$ (discussed in the previous subsection), the equality holds if and only if $\alpha_k=\alpha k$, where $\alpha$ is a constant. This case leads to the stationary $k$-th current (see Eq. \eqref{currentk}) $\mathcal{J}^{(k)}(x)^*= \alpha D_kP^*$, which is independent of $x$.}

{\color{black}\section{Continuous-state systems}}

Our results can be extended to a continuous-step model where the system can jump to \textit{any} location according to a certain distribution. The continuous versions of Eqs (\ref{discME}) and (\ref{schn}) are \cite{gardiner}:
\small
\begin{equation}
\dot{P}(x,t) = \int dr \left( W(x-r,r) P(x-r,t) - W(x,-r) P(x,t) \right)
\end{equation}
\normalsize
and
\begin{eqnarray}
\dot{S}_{\mathrm{ME}}(t) = \int &dx& \int dr W(x-r,r) P(x-r,t)  \nonumber \\
&\times& \log\frac{W(x-r,r)P(x-r,t)}{W(x,-r)P(x,t)}
\label{schncont}
\end{eqnarray}
where $W(x,r)$ is the rate density of a jump of size $r$ from location $x$. We now consider an infinite system and therefore integrals are performed between $-\infty$ and $+\infty$. 

We take the following scaling form for the transition rates:
\begin{equation}
W(x,r) = \frac{1}{\epsilon} \frac{1}{\sqrt{d(x) \epsilon}} e^{- f(z(x,r))},
\label{W}
\end{equation}
where $f$ is a generic symmetric function \cite{note2} and $z(x,r) = (r - A(x) \epsilon)/\sqrt{d(x) \epsilon}$. Without loss of generality we have chosen $f(0)$ such that $\int dz  e^{-f(z)} = 1$. We have introduced an expansion parameter $\epsilon$ in such a way to control the right scaling of $W(x,r)$ in the diffusive limit, $\epsilon \rightarrow 0$. The surviving terms in the Kramers-Moyal expansion lead to the FPE, Eq. \eqref{FPE}, with $D(x)=d(x)\int dz\, z^2 e^{-f(z)} /2$ {\color{black}(see Appendix A for details)}.

The entropy production, calculated in the  $\epsilon\rightarrow0$ limit, is {\color{black}(the derivation is quite lengthy and it is presented in Appendix B)}: 
\begin{eqnarray}
\dot{S}_{\mathrm{ME}}^{\epsilon\rightarrow 0}\equiv \lim_{\epsilon\rightarrow 0}\dot{S}_{\mathrm{ME}}(t) &=& \int dy \frac{\left( A(y) P(y) - \partial_y \left( D(y) P(y) \right)\right)^2}{D(y) P(y)} + \nonumber \\
&+& \left( \langle z^2 \rangle \langle (\partial_z f(z))^2 \rangle - 1 \right) \int dy \frac{A(y)^2}{D(y)} P(y) + \nonumber \\
&+& \left( 3 - \langle z^2 (\partial_z f(z))^2 \rangle \right) \int dy \frac{A(y) \partial_y D(y)}{D(y)} P(y) + \nonumber \\
&+& \frac{1}{4} \left( - 9 + \frac{\langle z^4 (\partial_z f(z))^2 \rangle}{\langle z^2 \rangle} \right) \int dy \frac{(\partial_y D(y))^2}{D(y)} P(y)
\label{Entgen}
\end{eqnarray}
where $\langle\cdot\rangle = \int dz \cdot e^{-f(z)}$.
Since this general formula is quite cumbersome, in what follows we restrict our analysis to two simple cases of interest: the one with non-vanishing drift and constant diffusion rate, and the case with zero drift and space-dependent diffusion coefficient.

{\color{black}\subsection{Two simple frameworks and limit of no correction}}

For a constant diffusion coefficient ($D(x)=D$), we obtain:
\begin{equation}
\dot{S}_{\mathrm{ME}}^{\epsilon\rightarrow0} = \dot{S}_{\mathrm{FP}} + \left( \langle z^2 \rangle \langle \partial_z f(z) \rangle -1 \right) \int dx \frac{A(x)^2}{D} \geq \dot{S}_{\mathrm{FP}},
\label{ineq1}
\end{equation} 
where the inequality follows from the Cauchy-Schwarz inequality. Eq. (\ref{ineq1}) emphasizes that  Seifert's formula \eqref{seifert} needs to be corrected by a positive term, which takes into account information about the microscopic dynamics missing in the FPE.

It is particularly interesting the choice of Gaussian transition rates, $f(z) = z^2 + \log\sqrt{\pi}$. This represents the limiting case, in this setting of constant diffusion, where there is no loss of information in the coarse-graining process, so that Eq. (\ref{ineq1}) holds as equality{\color{black}, i.e. the Seifert's formula gives the actual entropy production}. This result agrees with the fact that, in order to consistently describe a microscopic dynamics as a FPE, one needs to assume Gaussian transition rates, otherwise inconsistencies in non-equilibrium quantities may arise \cite{mazur, mazur2}.

{\color{black} Notice that, in principle, there could be physical systems exhibiting non-Gaussian transition rates. It can be seen, for some cases {\color{black}(see Appendix C)}, that this rely on how the energy barriers between any two states behave as a function of their distance (in a real or abstract state space).}

On the other hand, when $A=0$ and $D(x)$ is not constant, we obtain:
\begin{equation}
\dot{S}_{\mathrm{ME}}^{\epsilon\rightarrow0} = \int dx \frac{J(x)^2}{D(x) P(x)} + \gamma \int dx \frac{(\partial_x D(x))^2}{D(x)},
\label{ineq2}
\end{equation}
%\end{widetext}
where $\gamma=  \left( - 9 + \langle z^4 (\partial_z f(z))^2\rangle/\langle z^2 \rangle \right)/4$. {\color{black} To demonstrate the positivity of $\gamma$, let us define a new measure (not a probability measure) $d\mu = dz z^2 e^{-f(z)}$, then:
\begin{equation}
\frac{\langle z^4 (\partial_z f(z))^2 \rangle}{\langle z^2 \rangle} = \frac{\langle z^2 (\partial_z f(z))^2 \rangle_{\mu}}{\langle 1 \rangle_{\mu}}
\end{equation}
where:
\begin{equation}
\langle \; \cdot \; \rangle_{\mu} = \int \cdot \; d\mu = \langle \; \cdot \; z^2 \rangle
\end{equation}
Noting that:
\begin{equation}
\langle z^2 (\partial_z f(z))^2 \rangle_{\mu} \langle 1 \rangle_{\mu} \geq \langle z \partial_z f(z) \rangle_{\mu}^2
\end{equation}
the inequality directly follows by rearrangement and integration by parts. Note that, in this case, the inequality holds for any choice of $W(y,r)$, which is consistent with the Kramers-Moyal expansion.}
Although this result is valid for any system amenable to be described by a ME, it is interesting to study the application to the case of a diffusing particle. The entropy production in Eq. \eqref{ineq2} corresponds to the one associated with the Fokker-Planck description of an overdamped process, where the  positive corrections are due to the coarse-graining procedure as explained above.

{\color{black}\subsection{Future perspectives}}

{\color{black}When the Fokker-Planck Equation exhibits a} non-constant diffusion coefficient, $D(x)$, the underdamped setting represents a more appropriate description of the system. The generalization of our result to this case will be investigated in future works. It is important to say that the discrepancy between the entropy production in the underdamped and overdamped regime, investigated within the formalism of the FPE, as in \cite{pre,celani}, comes from a different source and does not involve coarse-graining corrections nor information about the microscopic transition rates.

{\color{black}Experimental analysis based on the theory here presented would be useful and interesting, in particular to acquire some information about the process underlying a diffusive description. In fact,} our approach relies on knowing many microscopic details of the system --the transition rates-- which are commonly unknown or not properly measurable. A simple experimental setup could be provided by a one-dimensional overdamped colloidal particle with a space-dependent diffusion and zero drift, similar to the one described in \cite{celani}. In this simple scenario, the corrections to the entropy production given by Eq. (\ref{ineq2}) do not vanish (even the simplest Gaussian case $f = z^2 + \log\sqrt{\pi}$ leads to $\gamma=3/2$, and therefore might become quantifiable by an experimental test.
\\

{\color{black}\section{Conclusions and open questions}}

It is well-known that a coarse-graining procedure {\color{black}leads to an underestimation of the entropy production \cite{espos, ziener, bo}. We have proven that the same applies when a mesoscopic description of the dynamics is adopted, i.e. when a coarse-graining is performed on the dynamics. {\color{black}In other words, within some limiting procedure, a dynamics described by a FPE can be derived from a microscopic ME and the two can be considered equivalent to many extents. However, we have shown that, in general, the entire non-equilibrium behaviour, as manifested, for example, in the entropy production,} cannot be predicted correctly by the FPE alone, as some further important information survive in the limiting procedure of the dynamics.}

{\color{black} On this regard a future perspective would be to look for a ``modified" FPE, which is able to capture all the relevant information surviving the coarse graining limiting procedure, rather than searching for a ME leading to the same entropy production as predicted by the standard FPE. This would result in a deeper understanding of the microscopic world hidden behind a coarse-grained description. It would also have consequences in fields ranging from artificial molecular motors \cite{raz,busie2,barato} to the possible quantification of the dissipation in biological systems \cite{horow} through the celebrated thermodynamic uncertainty relations \cite{dechant}. Furthermore, our results can be applied to cases where an evolution occurs over a generic state space such as in interacting ecological and social systems (e.g. bacteria, species, humans)  \cite{amos,mckane}. In this latter context, {\color{black}the description both in terms of both a ME and as a diffusive process (FPE) is usually known.} Thus, an information-theoretic and thermodynamic approach to population dynamics could lead to a better evaluation of the non-equilibrium activity and to a better identification of the relevant physical quantities in play.}

{\color{black}We would like to stress that the work presented here has been focused on the derivation of the corrections to the average entropy production under coarse-graining. In the field of stochastic thermodynamics, however, it is crucial to study quantities at the trajectory level. How to define a coarse-graining procedure that works on trajectories is a fundamental open question, whose answer could shed some light on the way to lose the least amount of information in describing a physical system.}
\\

\ack{
We acknowledge C. Jarzynski, C. Maes, U. Seifert and S. Suweis for useful comments and discussions. {\color{black}We also thank the anonymous reviewers for giving us ideas to improve the impact of our work. A.M. was supported by 'Excellence Project 2017' of the Cariparo Foundation.}
}

{\color{black}
\appendix

\section{Transition rates and FPE coefficients}

We derive the drift and diffusion coefficients for the following general choice for the transition rates:

\begin{equation}
W(y,r) = \frac{1}{\epsilon} \frac{1}{\sqrt{\epsilon d(y)}} e^{-f\left(\frac{r - A(y) \epsilon}{\sqrt{\epsilon d(y)}} \right)}
\end{equation}

where $f$ is a generic symmetric function. All the ``pseudo-moments" can be computed as follows:

\begin{eqnarray*}
a^{(n)} &=& \int r^{n} W(y,r) dr \rightarrow z = \frac{r-A(y) \epsilon}{\sqrt{d(y) \epsilon}} \rightarrow \nonumber \\
&\to& \int dz \sum_{k=0}^n  {n \choose k} z^{n-k} e^{-f(z)} A(y)^{k} d(y)^{\frac{n-k}{2}} \epsilon^{\frac{n+k}{2}-1}= \nonumber \\ 
&=& \sum_{k=0}^n {n \choose k} <z^{n-k}> A(y)^{k} d(y)^{\frac{n-k}{2}} \epsilon^{\frac{n+k}{2}-1}
\label{an}
\end{eqnarray*}
where:
\begin{equation}
\langle z^{n} \rangle = \int z^{n} e^{-f(z)} dz
\end{equation}

Up to the leading order in $\epsilon$, we get:

\begin{eqnarray*}
&\;\;\; a^{(1)} = \langle 1 \rangle A(y) \nonumber \\
&\;\;\; a^{(2)} = \langle z^2 \rangle d(y) + \langle 1 \rangle A(y)^2 \epsilon = \langle z^2 \rangle d(y) \nonumber \\
&a^{(n>2)} = \mathcal{O}\left( \epsilon^{\frac{n}{2}-1} \right) = 0
\label{aneps}
\end{eqnarray*}

where $\langle 1 \rangle$ is just the normalization of the transition rates. Then the Kramers-Moyal expansion can be performed, leading to a consistent FPE with a drift $A(y) \langle 1 \rangle$ and a diffusion coefficient $D(y) = \langle z^2 \rangle d(y)/2$. Let us define, for sake of simplicity, the following rescaled average:

\begin{equation}
\langle \cdot \rangle_0 = \frac{\langle \cdot \rangle}{\langle 1 \rangle}
\end{equation}

In what follows we will set $\langle 1 \rangle$ to be generic, even though it is possible to see that, without loss of generality, we can choose $f(z)$ such that $\langle 1 \rangle = 1$, as in the main text. 

\section{Splitting the entropy production}

The formula for the entropy production derived performing the diffusive limit on the Schnakenberg's expression can be rewritten as follows:

\begin{eqnarray*}
\dot{S} &=& \int \Bigg( \boldsymbol{s_1(y)} \frac{(\partial_y P(y,t))^2}{P(y,t)} + \boldsymbol{s_6(y)} \partial_y^2 P(y,t) + \nonumber \\
&+& \boldsymbol{\left( s_2(y) + \partial_y s_4(y) + \partial_y a^{(1)}(y) + s_5(y) + \partial_y s_7(y) + s_8(y) \right)} P(y,t) + \nonumber \\
&+& \boldsymbol{\left( s_3(y) + s_4(y) + s_7(y) + \partial_y s_6(y) \right)} \partial_y P(y,t) \Bigg) dr
\end{eqnarray*}

where:

\begin{eqnarray*}
{\color{black}s_1(y)} &=& \int \frac{r^2}{2} W(y,r) dr \nonumber \\
{\color{black}s_2(y)} &=& \int \frac{r^2}{2} \frac{(\partial_y W(y,-r))(\partial_y W(y,r))}{W(y,r)} dr \nonumber \\
{\color{black}s_3(y)} &=& \int \frac{r^2}{2} \left(\frac{W(y,-r)}{W(y,r)} + 1\right) \partial_y W(y,r) dr \nonumber \\
{\color{black}s_4(y)} &=& - \int \frac{r}{2} \left( W(y,r) + W(y,-r) \right) \log\left( \frac{W(y,r)}{W(y,-r)} \right) dr \nonumber \\
{\color{black}s_5(y)} &=& \int r \frac{W(y,-r)}{W(y,r)} \partial_y W(y,r) dr \nonumber \\
{\color{black}s_6(y)} &=& \int \frac{r^2}{2} (W(y,r) - W(y,-r)) \log \left( \frac{W(y,r)}{W(y,-r)} \right) dr \nonumber \\
{\color{black}s_7(y)} &=& \int \frac{r^2}{2} \partial_y (W(y,r) - W(y,-r)) \log \left( \frac{W(y,r)}{W(y,-r)} \right) dr \nonumber \\
{\color{black}s_8(y)} &=& \int \frac{1}{2} (W(y,r) - W(y,-r)) \log\left( \frac{W(y,r)}{W(y,-r)} \right) dr \nonumber
\end{eqnarray*}

{\color{black} with the implicit assumption that $P(x)$ vanishes, along with its derivative, at the boundaries.}

In what follows we will explicit the proposed form for the transition rates deriving an expilict expression for each one of these terms as function of the mescoscopic parameters $A(y)$ and $D(y)$ only.
\\

\subsection*{Expansion in $\epsilon$}

Here we introduce some useful expansions:

\textcolor{black}{
\begin{eqnarray}
\;\;\;\;\;\;\;\;\;\;\;\;\;\;\;\;\;\;\; z(y) &= &\frac{r}{\sqrt{\epsilon d(y)}} + \mathcal{O}(\sqrt{\epsilon}) \\
\;\;\;\;\;\;\;\;\;\;\;\;\;\;\;\; \partial_y z(y) &= &-\frac{1}{2} \frac{1}{\sqrt{\epsilon d(y)}} \frac{\partial_y d(y)}{d(y)} r + \mathcal{O}(\sqrt{\epsilon})\nonumber \\
f\left(\frac{- r - A(y) \epsilon}{\sqrt{\epsilon d(y)}}\right) &= &f(-z) - \sqrt{\epsilon}\frac{2 A(y)}{\sqrt{d(y)}} \partial_z f(-z) + \mathcal{O}(\epsilon)
\end{eqnarray}
}

\subsection*{Diffusive limit of the Schnakenberg's entropy production}

Reminding that $f\left( \frac{r-A(y) \epsilon}{\sqrt{\epsilon d(y)}} \right) \equiv f(z)$, using all the approximations introduced above and the parity of $f(z)$, up to the order $\mathcal{O}(1)$ in $\epsilon$, we get:

\begin{itemize}
\item
\begin{equation}
{\color{black} s_1(y)} = \int \frac{r^2}{2} W(y,r) dr = D(y)
\label{a(y)}
\end{equation}

\item 
\begin{eqnarray*}
{\color{black} s_2(y)} &=& \int \frac{r^2}{2} \frac{(\partial_y W(y,-r))(\partial_y W(y,r))}{W(y,r)} dr = \\ &=& \frac{1}{4} \frac{(\partial_y D(y))^2}{D(y)} \left(1+\frac{\langle z^4 (\partial_z f(z))^2 \rangle_0}{\langle z^2 \rangle_0} - 2 \frac{\langle z^3 \partial_z f(z) \rangle_0}{\langle z^2 \rangle_0} \right)
\end{eqnarray*}

\item
\begin{eqnarray*}
{\color{black} s_3(y)} = \int \frac{r^2}{2} \left(\frac{W(y,-r)}{W(y,r)} + 1\right) \partial_y W(y,r) dr = 2 \partial_y D(y)
\end{eqnarray*}

\item 
\begin{eqnarray*}
{\color{black} s_4(y)} = - \int \frac{r}{2} \left( W(y,r) + W(y,-r) \right) \log\left( \frac{W(y,r)}{W(y,-r)} \right) dr = -2 A(y)
\end{eqnarray*}

\item 
\begin{eqnarray*}
{\color{black} s_5(y)} = \int r \left(\frac{W(y,-r)}{W(y,r)}\right) \partial_y W(y,r) dr = \partial_y A(y) - \frac{A(y) \partial_y D(y)}{D(y)} \left(\langle z^2 (\partial_z f(z))^2 \rangle_0 - 1 \right)
\end{eqnarray*}

\item
\begin{eqnarray*}
{\color{black} s_6(y)} = \int \frac{r^2}{2} (W(y,r) - W(y,-r)) \log \left( \frac{W(y,r)}{W(y,-r)} \right) dr = 0
\end{eqnarray*}
 
\item 
\begin{eqnarray*}
{\color{black} s_7(y)} = \int \frac{r^2}{2} \partial_y (W(y,r) - W(y,-r)) \log \left( \frac{W(y,r)}{W(y,-r)} \right) dr = 0
\end{eqnarray*}

\item
\begin{eqnarray*}
{\color{black} s_8(y)} &=& \int \frac{1}{2} (W(y,r) - W(y,-r)) \log \left( \frac{W(y,r)}{W(y,-r)} \right) dr = {\color{black}\frac{A(y)^2}{D(y)} \langle z^2 \rangle_0 \langle \left(\partial_z f(z)\right)^2 \rangle_0}
\label{dytot}
\end{eqnarray*}
\end{itemize}

Putting all the terms together:

\begin{eqnarray*}
\dot{S} &=& \int dy \frac{J(y)^2}{D(y) P(y)} + \left( \langle z^2 \rangle_0 \langle (\partial_z f(z))^2 \rangle_0 - 1 \right) \int dy \frac{A(y)^2}{D(y)} P(y) + \nonumber \\
&\;& + \left( 3 - \langle z^2 (\partial_z f(z))^2 \rangle_0 \right) \int dy \frac{A(y) \partial_y D(y)}{D(y)} P(y) \\
&\;& + \frac{1}{4} \left( - 9 + \frac{\langle z^4 (\partial_z f(z))^2 \rangle_0}{\langle z^2 \rangle_0} \right) \int dy \frac{(\partial_y D(y))^2}{D(y)} P(y)
\label{sgen}
\end{eqnarray*}

\section{Non-Gaussian transition rates}

Non-Gaussian transition rates might arise in systems, whose states are denote by $i$,  with a complex free energy landscape, $E_i$, where the Arrhenius form for the transition rates holds:
\begin{equation}
W_{ij} = e^{-\frac{B_{ij}-E_j}{\kappa_BT}}
\end{equation}
for all the off-diagonal elements, while $W_{ii} = -\sum_{k\neq i} W_{ki}$. If the system eventually relaxes to an equilibrium point, $B_{ij} = B_{ji}$. $B_{ij}$ can be interpreted as an effective free energy barrier (activation energy) between the state $i$ and $j$. For some applications of this form of the transition matrix, see \cite{raz1,jarz1,astum1}.

It is then easy to see that, if $B_{ij} = E_j + f(i,j)$, the function $f$ defines the behaviour of the transition rates as the effective distance between states increases. Then, Gaussian transition rates would occur for very special choices of the function $f$.
}


\begin{thebibliography}{54}
\bibitem{note}
\bibinfo{note}{We called them ``pseudo-moments", since $W(y,r)$ is not a
  distribution.}
\bibitem{notefoot1}
\bibinfo{note}{{\color{black}We alert the reader that the continuum limit in this example has been carried out naively. We postpone a more formal derivation to the general case of a continuous state space.}}
\bibitem{notefoot2}
\bibinfo{note}{{\color{black}The generalization to $A$ and $D$ depending on the state of the system is straightforward.}}
\bibitem{note2}
\bibinfo{note}{The symmetry condition on $f$ ensures that $\langle z \rangle =
  0$. This condition is necessary to have {\color{black}the drift coefficient of
  order $1$ (see Appendix A for details).}}
\end{thebibliography}
\end{document}